\documentclass[preprint,number,sort&compress]{elsarticle}
\usepackage{amsmath}
\usepackage{amssymb}
\usepackage[labelfont=bf,labelsep=period]{caption}
\usepackage{makecell}

\biboptions{comma,square}

\def\fref#1{{Fig. \ref{#1}}}
\def\tref#1{{Table \ref{#1}}}

\journal{Chemical Physics}

\begin{document}

\begin{frontmatter}

\title{Time of Flight Transients in the Dipolar Glass Model}
\author[label1]{S.V. Novikov\corref{cor}}
\ead{novikov@elchem.ac.ru}
\cortext[cor]{Corresponding author}
\author[label2]{A.P. Tyutnev}
\author[label3]{L.B. Schein\fnref{label_d}}
\address[label1]{A.N. Frumkin Institute of
Physical Chemistry and Electrochemistry, Leninsky prosp. 31, Moscow 119991, Russia}
\address[label2]{Moscow State Institute of Electronics and Mathematics, Bol. Trechsvyatitel. per., 3,
Moscow 109028, Russia}
\address[label3]{Independent Consultant, 7026 Calcaterra Drive, San Jose, California 95120, USA}
\fntext[label_d]{Deceased}


\begin{abstract}
Using Monte Carlo simulation we investigated time of flight current transients predicted by the dipolar glass model for a random spatial distribution of hopping centers. Behavior of the carrier drift mobility was studied at room temperature over a broad range of  electric field and sample thickness. A flat plateau followed by $j\propto t^{-2}$ current decay is the most common feature of the simulated transients. Poole-Frenkel mobility field dependence was confirmed over 5 to 200 V/$\mu$m as well as its independence of the sample thickness. Universality of transients with respect to both field and sample thickness has been observed. A simple phenomenological model to describe simulated current transients has been proposed. Simulation results agree well with the reported Poole-Frenkel slope and shape of the transients for a prototype molecularly doped polymer.
\end{abstract}


\end{frontmatter}



\section{Introduction}

Dipolar glass (DG) model \cite{Novikov:14573,Dunlap:542,Novikov:4472,Dunlap:437,Parris:126601} was developed in the late 1990s in response to the urgent need to explain the ubiquitous Poole-Frenkel (PF) mobility field dependence observed in amorphous polymers (polyvinylcarbazole \cite{Gill:5033}, polysilenes \cite{Abkowitz:817} and others), molecularly doped polymers \cite{Schein:4287,Schein:686,Borsenberger:5188,Borsenberger:6263}, and low molecular weight organic glasses \cite{Borsenberger:465} in a broad field range. This model overcomes the limitations of the Gaussian Disorder Model (GDM) \cite{Bassler:15}, which successfully explained the mobility temperature dependence, but failed to reproduce the mobility field dependence for small and moderate electric fields. The DG model is a natural evolution of the earlier approach of Borsenberger and Bassler \cite{Borsenberger:5327,Borsenberger:185}, who suggested dipolar energetic disorder as an explanation of strong polarity effect on the hopping charge mobility in amorphous organic materials. They considered the dipolar as well van der Waals energetic disorder, but did not  introduce them consistently on a microscopic level. It turned out that strong spatial correlations in the random energy landscape, intrinsic to dipolar disorder, move transport properties of polar materials far away from those of the GDM. In particular, spatial correlations naturally provide a strong physical foundation for the development of the PF mobility field dependence \cite{Dunlap:542,Novikov:4472}.

Previous Monte Carlo (MC) simulations \cite{Novikov:444} have shown that the time of flight (TOF) currents, predicted by the DG model, provide transients having typical features of the experimental ones: an initial short spike, reflecting a non-equilibrium stage of the carrier transport, followed by a flat plateau signaling partial equilibration of carriers, and then an anomalously long tail described by a power law  $j\propto t^{-\beta}$ ($\beta\approx  2.0-2.5$). All earlier models failed to predict these broad post flight current tails in combination with almost flat plateau for the case of strong disorder.

The aim of the present paper is to investigate in detail the TOF currents, predicted by the DG model for a prototype molecularly doped polymer 30\% DEH:PC (polycarbonate doped with 30 wt.\% of aromatic hydrazone DEH).

\section{Basics}

We consider charge transport in a cubic lattice of randomly oriented static dipoles with the dipole moment  $p$, where $c$ is the fraction of sites serving as hopping centers (transport sites). The random energy of a particular transport site is $U_i = e\varphi(\vec{r}_i)$, where $\varphi(\vec{r}_i)$ is the electrostatic potential, created by all other dipoles. If $c$ is not too low, then the density of states has a Gaussian form and correlation function $C(\vec{r})=\left<U(\vec{r})U(0)\right>$ decays as $1/r$ \cite{Novikov:14573,Dunlap:542,Dieckmann:8136}. If electric field $F$ is applied, then the random energy has an additional term $-e\vec{F}\vec{r}_i$. The DG model is also known as a Correlated Disorder Model (CDM) \cite{Novikov:4472}, but the present name is better suited for description of the true nature of the model; for example, it explicitly assumes a particular spatial decay of the correlation function $C(\vec{r})\propto 1/r$, while the name "CDM" is too ambiguous in this respect. Indeed, the model of the quadrupolar glass, where randomly oriented quadrupoles fill the lattice instead of dipoles, provides a correlated energy landscape as well, so it again can be dubbed a Correlated Disorder Model, but the correlation function in the quadrupolar case is different, $C(\vec{r})\propto 1/r^3$ \cite{Novikov:181,Novikov:2584}, and, hence, the transport properties are different as well \cite{Novikov:954}.

The Miller-Abrahams hopping rate was used for the simulations, where the rate of transition from site $i$ to site $j$ is given by
\begin{equation}\label{MA_rate}
p_{i\rightarrow j}=\nu_0 \exp(-2\gamma r_{ij})\begin{cases}
    \exp\left(-\frac{U_j-U_i}{kT}\right), &U_j-U_i > 0\\1, &U_j-U_i < 0
\end{cases}
\end{equation}
here $\nu_0$ is the prefactor frequency, $r_{ij}=|\vec{r}_j-\vec{r}_i|$, and $\gamma$ is a wave function decay parameter for transport sites.

Monte Carlo simulations refer to the traditional TOF  experiment when a sheet of carriers (holes in the case of DEH:PC polymer) is instantly produced at the generating electrode at $x=0$ and then drifts towards the collecting electrode located at $x=L$  (hence, $L$ is a thickness of the transport layer). The output of the simulation is the mean velocity $v(t)$ of a carrier; the major experimental observable, current density $j(t)$, is proportional to  $v(t)$. Another important parameter is the mean carrier velocity $\left<v\right>$ for a full transfer from the generating to collecting electrode. This velocity is equal to $\left<L/t_{\rm d}\right>$, where   $t_{\rm d}$ is the time for a carrier to drift from $x= 0$ to $x= L$. Other details of the simulation can be found elsewhere \cite{Novikov:4472,Bassler:15}.

In the simulation we tried to model the polycarbonate doped with 30 wt.\% of aromatic hydrazone DEH. DG model parameters (borrowed from the analysis of experimental data in Ref. \cite{Schein:1067}) are as follows: the rms dipolar disorder $\sigma$  is 0.13 eV and contains no additional contributions (e.g., van der Waals), the lattice constant $a = 0.77$ nm,  $2\gamma a = 11.8$, $kT = 0.0252$ eV (room temperature), and $c = 0.3$.

In all figures (apart from \fref{fig10}) we use a dimensionless time $t=\nu_0\exp\left(-2\gamma a\right) t'$,  where $t'$ is the physical time. Frequency $\nu_0$ was taken to be $1\times 10^{16}$ s$^{-1}$ to provide matching of the simulated carrier velocity at 5 V/$\mu$m with the experimental one from Ref. \cite{Schein:1067}.

\section{Simulation results}

\begin{figure}[tbh]
\begin{center}
\includegraphics[width=3in]{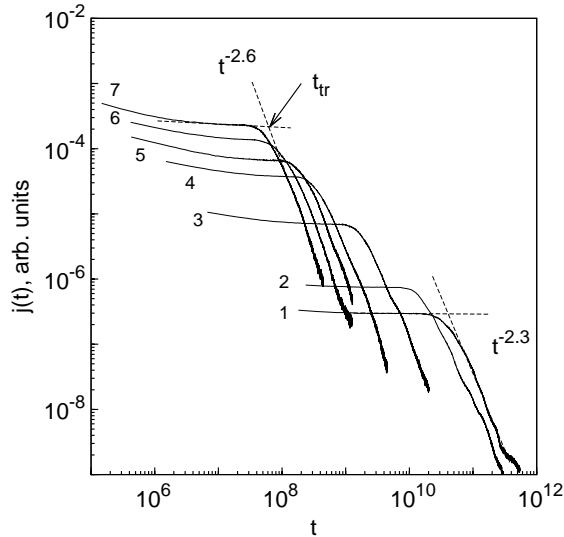}
\end{center}
\caption{Simulated TOF transients for the electric field equal to 5 (1), 10 (2), 36 (3), 80 (4), 110 (5), 150 (6), and 200 V/$\mu$m (7), correspondingly; $L=20\thinspace000$ lattice planes. Time of flight $t_{tr}$ is shown by the arrow.\label{fig1}}
\end{figure}

\fref{fig1} shows TOF transients for  $L= 15.4$ $\mu$m ($20\thinspace000$ lattice planes) for fields in the range $5-200$ V/$\mu$m. In all cases the transport occurs in a quasi-stationary regime with well-defined transient plateaus. Log-log representation shows that the post flight current decay follows a power law $j\propto t^{-\beta_2}$ rather than an exponential one, usually expected for fully equilibrated transport with well-defined plateau of the transient. Time of flight $t_{tr}$ is determined as the intersection of the tangents to the initial (slope $\beta_1$) and post flight (slope $\beta_2$) parts of the curve as shown by broken straight lines in the figure. While $\beta_1$ is close to zero, parameter $\beta_2$ varies between 2.1 and 2.7. For fields exceeding 36 V/$\mu$m the initial current spike begins to appear indicating a non-equilibrium stage of the transport process.

\begin{figure}[tbh]
\begin{center}
\includegraphics[width=3in]{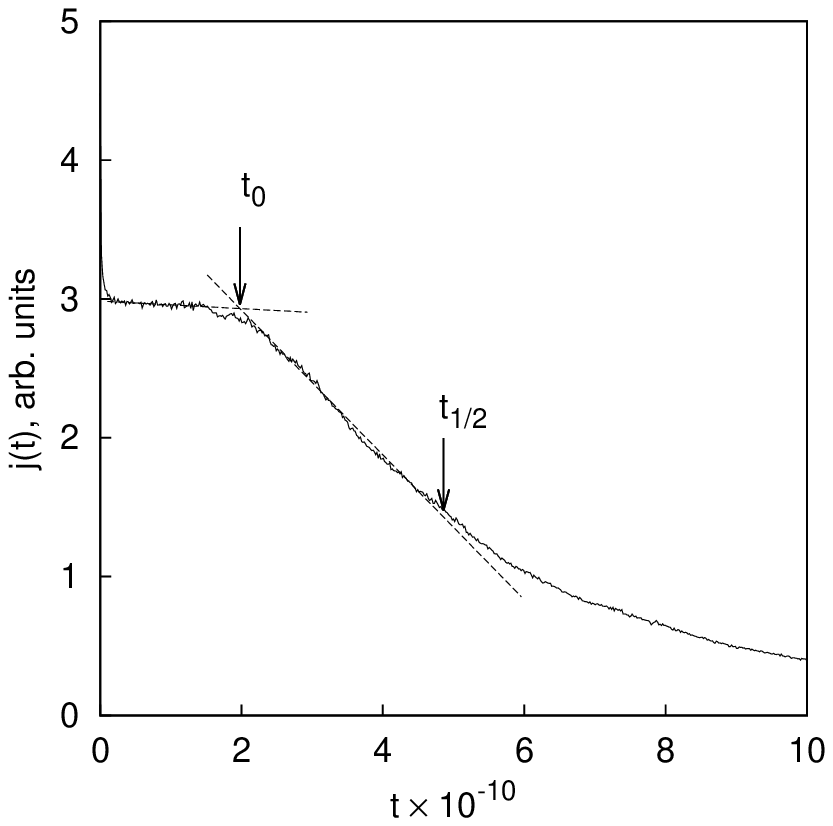}
\end{center}
\caption{Transient 1 from \fref{fig1} ($F = 5$ V/$\mu$m,   $L=20\thinspace000$ lattice planes) in linear coordinates with $t_0$ and $t_{1/2}$ indicated by arrows.\label{fig2}}
\end{figure}

Current transients can be presented in double linear coordinates to give traditional transit times $t_0$  or $t_{1/2}$ (shown in \fref{fig2}). Parameter $W=\left(t_{1/2}-t_0\right)/t_{1/2}$  is approximately constant in the range $5-200$ V/$\mu$m, fluctuating around 0.45.

\begin{figure}[tbh]
\begin{center}
\includegraphics[width=3in]{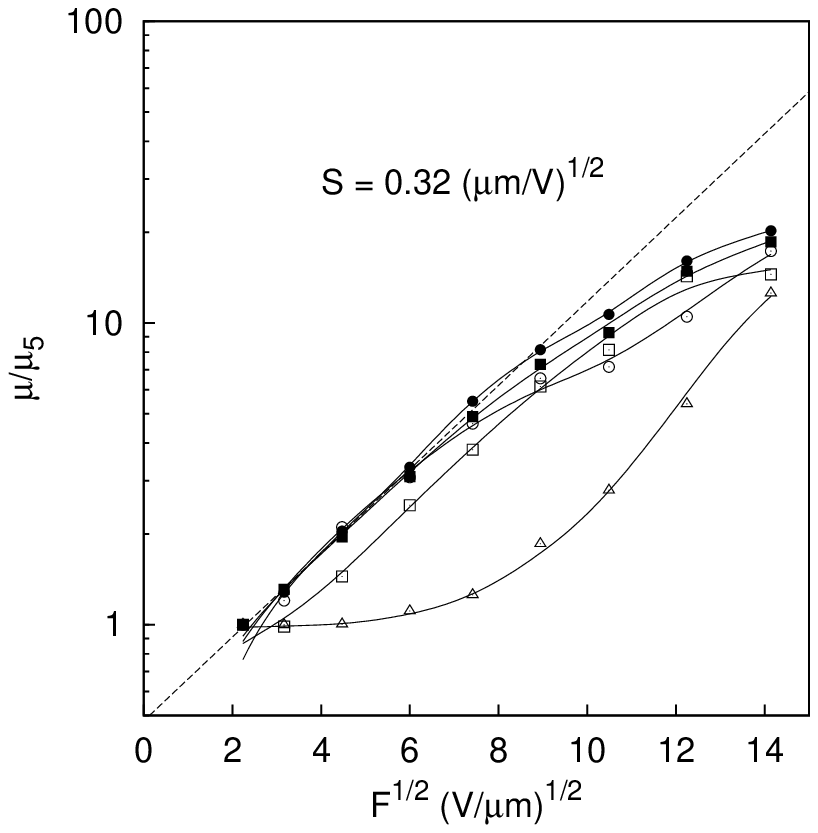}
\end{center}
\caption{Mobility field dependence (data points refer to the mobility defined by the different methods: using $t_0$ $(\Box)$, $t_{1/2}$ $(\blacksquare)$, $t_{tr}$ $(\circ)$, and $\left<v\right>$ $(\bullet)$, correspondingly ($\mu_5$ is the corresponding mobility for $F=5$ V/$\mu$m). Solid lines serve as a guide to the eye, and the broken line corresponds to the PF slope $0.32$ ($\mu$m/V)$^{1/2}$. The lowest curve $(\triangle)$ shows the GDM mobility, calculated using $\left<v\right>$.\label{fig3}}
\end{figure}

Transit times ($t_{tr}$, $t_0$, and $t_{1/2}$) for various fields are summarized in \tref{table1}. One can see that $t_{tr}$ is bounded by $t_{1/2}$ (upper bound) and  $t_0$ (within uncertainties in $t_{tr}$ due to errors in plotting the asymptote to the tail of the transient). \fref{fig3} shows field dependence of the relative drift mobility $\mu  = L/Ft$, calculated in four possible ways, using the corresponding values of $t_{tr}$,  $t_0$, $t_{1/2}$, as well as MC drift time $L/\left<v\right>$  (shown is the ratio of the mobility to the corresponding mobility at 5 V/$\mu$m). For low and moderate  fields all four curves are rather similar, they generally follow PF dependence
\begin{equation}
\mu_0\propto\exp\left(SF^{1/2}\right)
\label{PF}
\end{equation}
where $S$ is the PF slope. At high fields the calculated mobility field dependence begins to deviate from (\ref{PF}), reflecting the particular property of the Miller-Abrahams hopping rate \cite{Novikov:4472}.

\begin{table}[tbh]
  \centering
  \caption{Field dependence of transit times for $L=20\thinspace000$ lattice planes.\label{table1}}
\begin{tabular}{|c|c|c|c|}
\hline
\gape[t]{$F$, V/$\mu$m} &\multicolumn{3}{|c|}{Time of flight}        \\  \cline{2-4}
&   \gape[t]{$10^{-8}t_{tr}$}  & $10^{-8}t_{
0}$  & $10^{-8}t_{1/2}$ \\ \hline
                    5 &  346&   197 &  490  \\ \hline
                   10 &  144&   100 &  187  \\ \hline
                   20 & 41.1&  34.1 & 62.7  \\ \hline
                   36 & 15.6&  11.0 &   22  \\ \hline
                   55 & 6.78&  4.71 &  9.1  \\ \hline
                   80 &  3.3&   2.0 &  4.2  \\ \hline
                  110 &  2.2&   1.1 &  2.4  \\ \hline
                  150 &  1.1&  0.46 &  1.1  \\ \hline
                  200 & 0.5 & 0.34  & 0.66 \\ \hline
\end{tabular}
\end{table}

\begin{table}[tbh]
  \centering
  \caption{Thickness dependence of transit times.\label{table2}}
\begin{tabular}{|c|c|c|c|c|}
\hline
\gape[t]{$F$, V/$\mu$m} & $L$, lattice planes & \multicolumn{3}{|c|}{Time of flight} \\ \cline{3-5}
&        &     \gape[t]{$10^{-8}t_{tr}$} & $10^{-8}t_0$ &  $10^{-8}t_{1/2}$ \\ \hline
&   $1\thinspace000$     &  0.0228 &      &       \\ \cline{2-5}
200    &  $50\thinspace000$ &  1.26 &  0.925 &   1.68 \\ \cline{2-5}
&    $100\thinspace000$ &   2.74 & 1.96 &  3.40 \\ \cline{2-5}
&  $200\thinspace000$ &    5.26 &  4.04 &   6.76 \\ \hline
&     500 &   9.05 & 5.00 & 12.10 \\ \cline{2-5}
5  &  $5\thinspace000$ & 85.1 & 47.1 &  119.0 \\ \cline{2-5}
&     $20\thinspace000$& 346 &  197 & 490\\ \hline
\end{tabular}
\end{table}

Thus, three main results of the previous simulations (fast establishment of the quasi-stationary transport regime as evidenced by the flat plateau, the power law decay of the tail of the transients, and PF mobility field dependence) have been confirmed.

\begin{figure}[tbh]
\begin{center}
\includegraphics[width=3in]{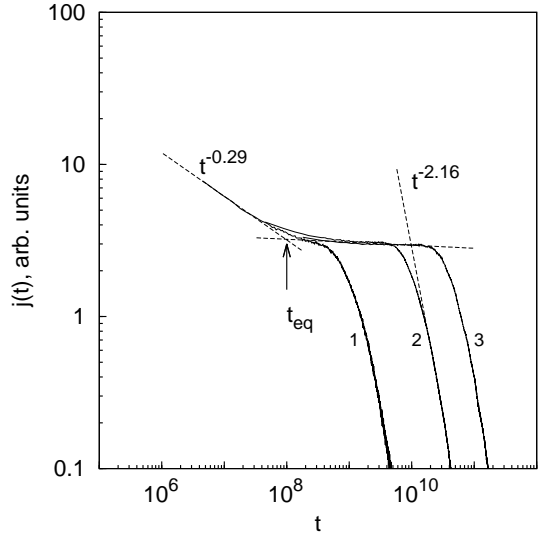}
\end{center}
\caption{Simulated TOF transients for  $L=$ 500 (1), $5\thinspace000$ (2), and $20\thinspace000$ (3) lattice planes, correspondingly. Electric field is 5 V/$\mu$m.\label{fig4}}
\end{figure}

\begin{figure}[tbh]
\begin{center}
\includegraphics[width=3in]{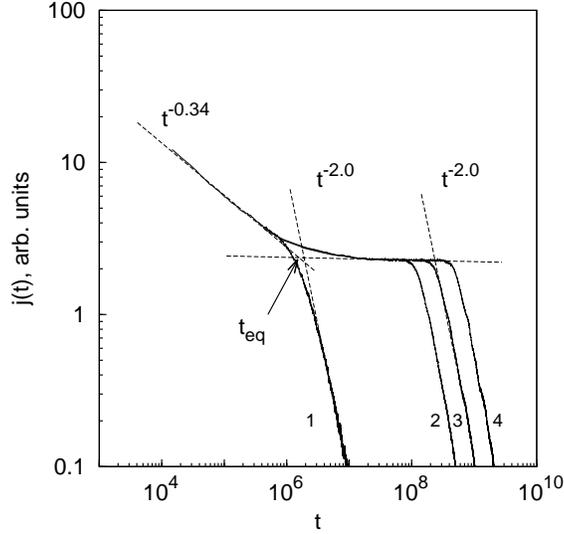}
\end{center}
\caption{Simulated TOF transients for $L$ equal to $1\thinspace000$ (1), $50\thinspace000$ (2), $100\thinspace000$ (3), and $200\thinspace000$ (4) lattice planes, correspondingly. Electric field is 200 V/$\mu$m.\label{fig5}}
\end{figure}

\begin{figure}[tbh]
\begin{center}
\includegraphics[width=3in]{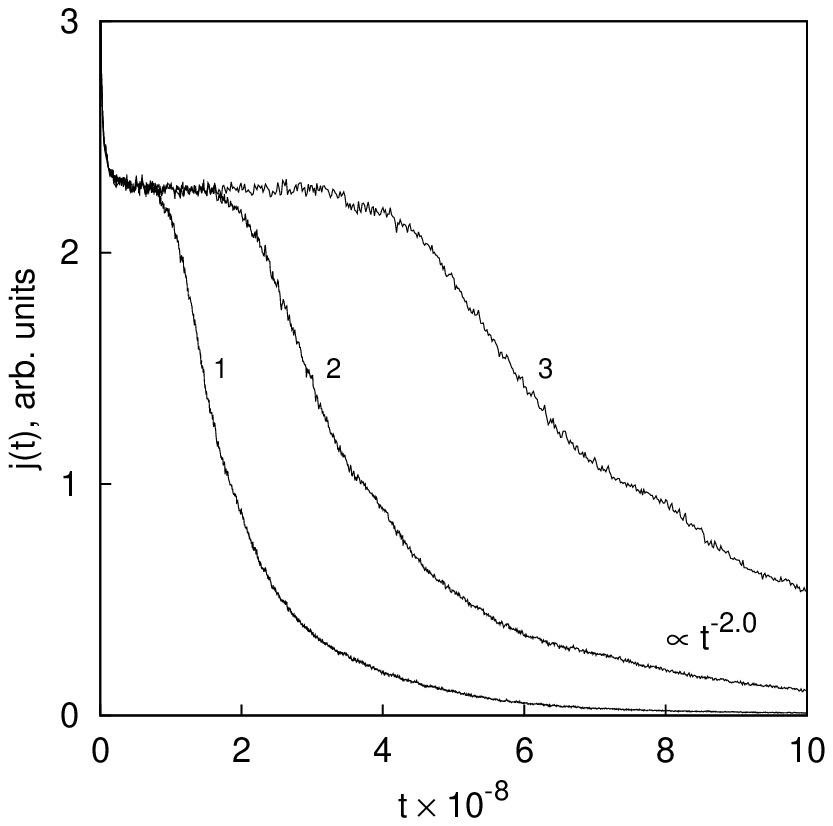}
\end{center}
\caption{Simulated TOF transients in linear representation for $L$ equal to $50\thinspace000$ (1), $100\thinspace000$ (2), and $200\thinspace000$ (3) lattice planes, correspondingly. These transients are the same as curves $2-4$ in \fref{fig5}. Electric field is 200 V/$\mu$m.\label{fig6}}
\end{figure}

A new observation concerns the carrier equilibration time $t_{eq}$, which is easily seen in \fref{fig1} at large electric fields but seems rather difficult to quantify as the non-equilibrium stage merges smoothly into the equilibrium one. To clarify this situation we present  \fref{fig4} -- \fref{fig6} showing TOF transients at extreme fields (5 and 200 V/$\mu$m, respectively) for varying thickness $L$. Equilibration time $t_{eq}$ can be estimated as the time of an intersection of the tangents to the initial non-equilibrium part of the transient and the plateau.

The use of this procedure gives the equilibration time $1.4\times 10^6$ for the highest field of 200 V/$\mu$m, while for the smallest field (5 V/$\mu$m) it is much larger and equal to $1.2\times 10^8$. The observed variation of $t_{eq}$ is even larger than that of the carrier mobility: the former rises by a factor of 80, while the latter increases by a factor of 20.

\begin{figure}[tbh]
\begin{center}
\includegraphics[width=3in]{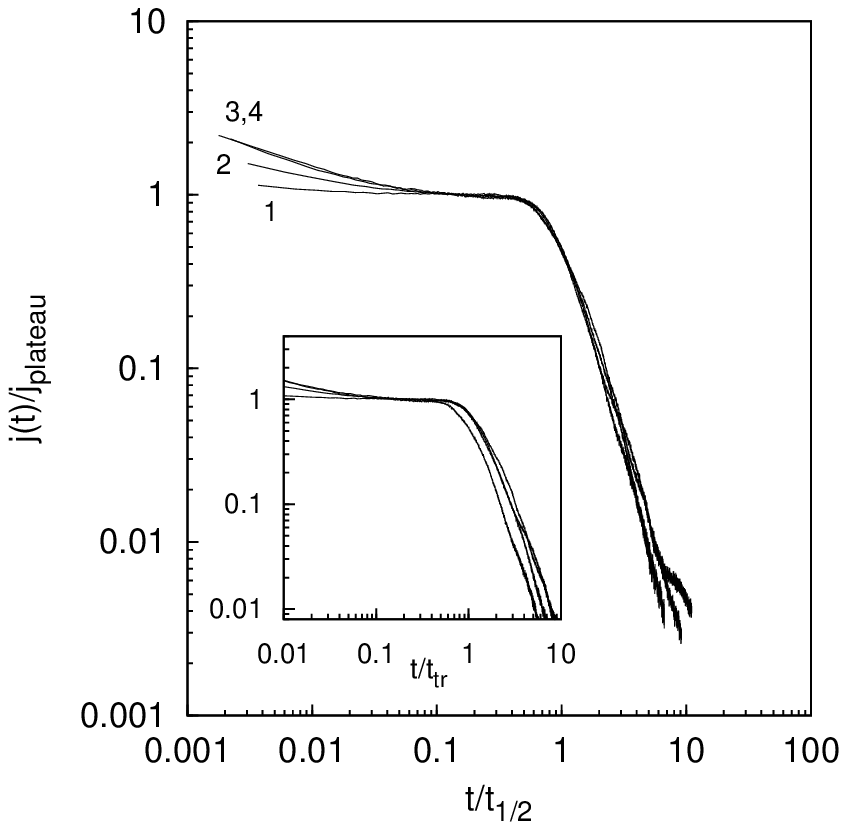}
\end{center}
\caption{Normalized transients for the electric field equal to 5 (1), 32 (2), 110 (3), and 200 V/$\mu$m (4), correspondingly; $L=20\thinspace000$ lattice planes. Curves 3 and 4 are almost indiscernible in this presentation.  Inset shows the analogous plot with $t_{tr}$ used as normalization time. Worse visual universality certainly follows from the much less reliable determination of $t_{tr}$ from log-log plots.\label{fig7}}
\end{figure}

\fref{fig1}, \fref{fig4}, and \fref{fig5} demonstrate features, usually attributed to the current universality traditionally observed in the continuous time random walk theory of Scher and Montroll \cite{Scher:2455} or multiple trapping formalism with the exponential trap distribution \cite{Rudenko:209}. Indeed, the pre-flight part of the TOF curves for the DG model is a flat plateau with $\beta_1\approx  0$, while the post flight decay again follows the power law dependence with the exponent  $\beta_2\approx 2.0$. These values agree well with the Scher-Montroll theory, where $\beta_1=1-\alpha$ and $\beta_2=1+\alpha$ with $\alpha \le 1$. Additionally, the transients demonstrate universality with respect to the field and thickness variation. The only contradicting factor is the existence of the equilibration time, which does not scale properly with the time of flight. For this reason we expect a mild violation of universality with respect to $F$ for high fields, as corroborated by \fref{fig7}.

\begin{figure}[tbh]
\begin{center}
\includegraphics[width=3in]{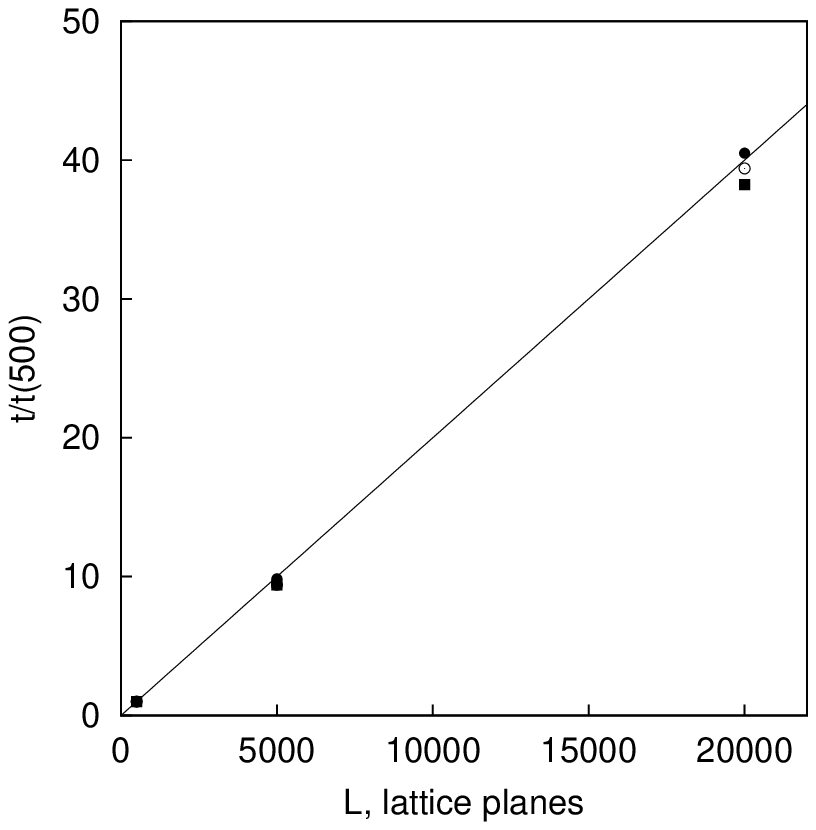}
\end{center}
\caption{Thickness dependence of time of flight for $t_{tr}$ $(\blacksquare)$, $t_0$ $(\circ)$, and $t_{1/2}$ $(\bullet)$, correspondingly; $t(500)$ is the corresponding time for $L=500a$ and straight line represents the linear dependence $t(L)=Lt(500)/500$. Electric field is 5 V/$\mu$m.\label{fig8}}
\end{figure}

Simulation data allows one to study the thickness dependence of the characteristic times of the transport process (\fref{fig4} -- \fref{fig6}). Obviously, $t_{eq}$ does not depend on $L$. Times of flight found on log-log as well as lin-lin plots exhibit thickness dependence very close to a linear one as could be seen from \fref{fig8} and \tref{table2} (for $F=200$ V/$\mu$m and $L=1000$ transient is too dispersive to calculate $t_0$ and $t_{1/2}$).

\begin{figure}[tbh]
\begin{center}
\includegraphics[width=3in]{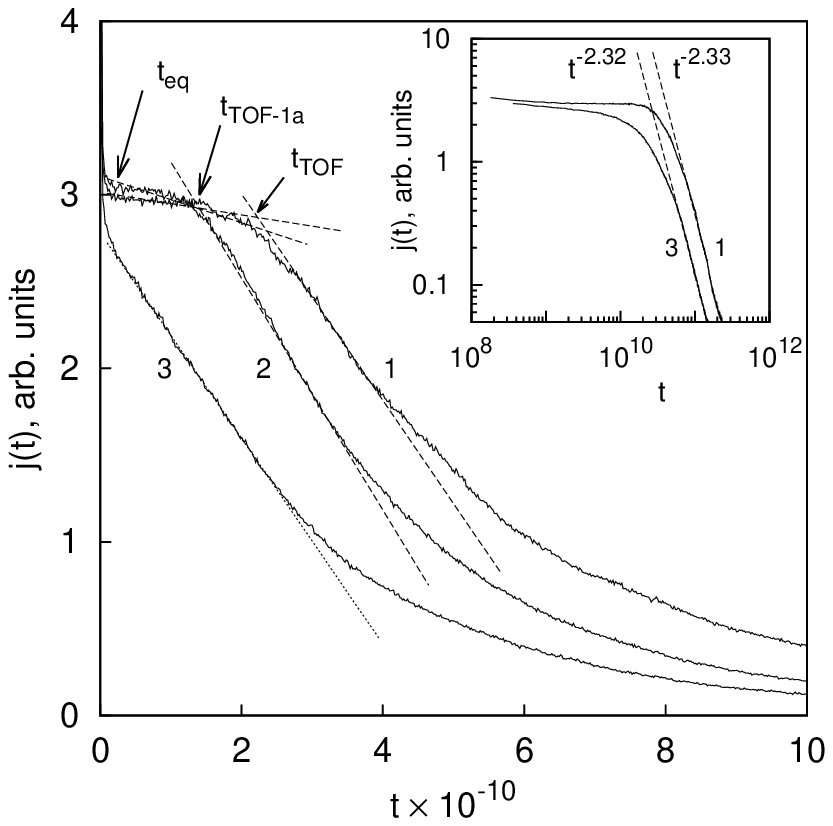}
\end{center}
\caption{Simulated TOF (1), TOF-1a (2) and TOF-2 (3) transients in linear coordinates. $L=20\thinspace000$ lattice planes, $F=5$ V/$\mu$m. Dotted straight line has been drawn to emphasize the linear decay of transient 3 in the time interval corresponding to the plateau of  transient 1; $t_{\rm TOF-1a}/t_{\rm TOF}\approx 0.63$. Inset shows TOF (1) and TOF-2 (3) transients in log-log plot. \label{fig9}}
\end{figure}

Until now we considered the classical TOF geometry where an instantaneous generation of the thin sheet of carriers takes place near the generating electrode. Recently two more geometries have been studied experimentally, namely, an instantaneous uniform generation of carriers in the bulk of the layer (TOF-2), or similar bulk carrier generation with a regulated depth of the generation zone (TOF-1a) \cite{Tyutnev:115107}. \fref{fig9} compares all three TOF variants for an equal total number of generated carriers (as always, in a small signal regime). As expected, if the TOF transient demonstrates a flat plateau, then the corresponding TOF-2 curve demonstrates an almost linear decay in the time range where the TOF current remains constant. Both transients decay similarly in the post-flight region, but the TOF current is always greater than the TOF-2 one, their ratio being practically constant (in our case it is equal to 3.2) as the log-log plot shows (\fref{fig9}, inset).

The TOF-1a transient, simulated for the thickness of the generation layer equal to $0.5L$,  demonstrates typical features of quasi-equilibrium transport. The plateau becomes more tilted and its length is approximately twice shorter (taking into account the accuracy of the MC simulation), as expected in this case \cite{Tyutnev:115107,Tyutnev:215219}.

\section{Discussion}

The most prominent features of the DG model are the PF field dependence of the carrier mobility, the particular shape of transients demonstrating the combination of a flat plateau with the slow post-flight current decay that follows a power law with the exponent $\beta_2$ close to 2.0, and current universality with respect to $F$ and $L$. The first feature was thoroughly discussed and explained earlier \cite{Dunlap:542,Novikov:4472,Dunlap:437}. Here we would like to discuss the specific behavior of transients. It is worth noting that universality of transients has already been reported in experimental papers for non-dispersive charge transport \cite{Schein:175,Borsenberger:967,Kreouzis:235201,Poplavskyy:415}.

Evidently, in our case development of a well-defined plateau cannot be described by the usual approach where the long time behavior is governed by the diffusion equation (such as the multiple-trapping model in the case of a fast decaying trap distribution \cite{Rudenko:163,Rudenko:177}) and current decay follows the exponential time dependence. An alternative explanation was suggested by Nikitenko et al. \cite{Nikitenko:136210}, who calculated an effective diffusion coefficient for the GDM and found that it slowly increases with time on a scale much longer that the typical equilibration time of the mean carrier velocity. Detailed analysis of the long time  behavior of transients was not performed but the particular case, shown in Figure 3 of Ref. \cite{Nikitenko:136210}, demonstrates approximate power law decay of the transient with the exponent $\beta_2$  close to 3. Applicability of this approach to the DG model is not evident, especially keeping in mind a well-established difference in the mobility field dependence in the GDM and DG model \cite{Novikov:4472,Novikov:2532}. In addition, MC simulations indicate that in the GDM for the case of strong disorder $\sigma/kT\simeq 5$  transients are much more dispersive: they do not develop a flat plateau and demonstrate a superlinear dependence of the transit time on the sample thickness \cite{Borsenberger:12145,Borsenberger:4289}. For these reasons we suggest a following simple phenomenological description of the charge transport.

Formally, transport behavior found in our simulation can be described by the "quasi-ballistic" model (qBM) where a distribution of drift carrier velocities takes place in a trap free organic solid. Indeed, let $f(v)$  be the density distribution of equilibrated carrier velocities. If a carrier moves with some velocity $v$, then at time  $t=L/v$ it reaches the collecting electrode and does not contribute to the total current anymore. Hence, in this model the current transient is
\begin{equation}
j(t)=en\int\limits_0^\infty dv v f(v)\theta\left(L-vt\right)
\label{qBM}
\end{equation}
where $\theta\left(x\right)$ is a unit step function, and $n$  is the initial density of carriers. Calculating the derivative, we obtain
\begin{equation}
\frac{dj}{dt}=-en\int\limits_0^\infty dv v^2 f(v)\delta\left(L-vt\right)=-en\frac{L^2}{t^3}f\left(L/t\right)
\label{qBM_2}
\end{equation}
or
\begin{equation}
f(v)=-\frac{L}{env^3}\left.\frac{dj}{dt}\right|_{t=L/v}.
\label{qBM_3}
\end{equation}
In general, if $j(t)\propto t^{-\beta_2}$  for $t\rightarrow \infty$, then $f(v)\propto v^{\beta_2-2}$  for $v\rightarrow 0$. Accordingly, for $\beta_2 = 2$ we have $f(v)\approx \textrm{const}$  for  $v\rightarrow 0$.

In this model TOF transients are universal and all transit times  are strictly proportional to $L$. Assuming an abrupt decay of $f(v)$  at $v=v_{\rm max}$ we automatically obtain a flat plateau for  $t\le L/v_{\rm max}$. Also, TOF-2 transient should have a linear decay extending for a full length of the TOF plateau, while the TOF-1a transient should have a shortened plateau in comparison to the original TOF one. Thus, all dynamic features predicted by the DG model for all three variants of the time of flight technique are reproduced by the model.

Limitations of the qBM are obvious: it cannot reproduce the non-equilibrium stage of the carrier transport, and the PF mobility field dependence has to be explicitly embedded into the model. Yet the real problem is to derive the phenomenological ballistic model directly from the basic hopping transport of the DG model.

At last we would like to discuss implications of the MC simulations regarding the prototype molecularly doped polymer 30\% DEH:PC thoroughly investigated in literature \cite{Schein:1067,Mack:7500,Tyutnev:115107}. At room temperature, it has been reported that PF slope is equal to 0.39 \cite{Schein:1067}, 0.42, \cite{Mack:7500} and 0.46 ($\mu$m/V)$^{1/2}$ \cite{Tyutnev:115107}. The PF slope for the straight line in \fref{fig3} is slightly smaller and equals 0.32 ($\mu$m/V)$^{1/2}$.

\begin{figure}[tbh]
\begin{center}
\includegraphics[width=3in]{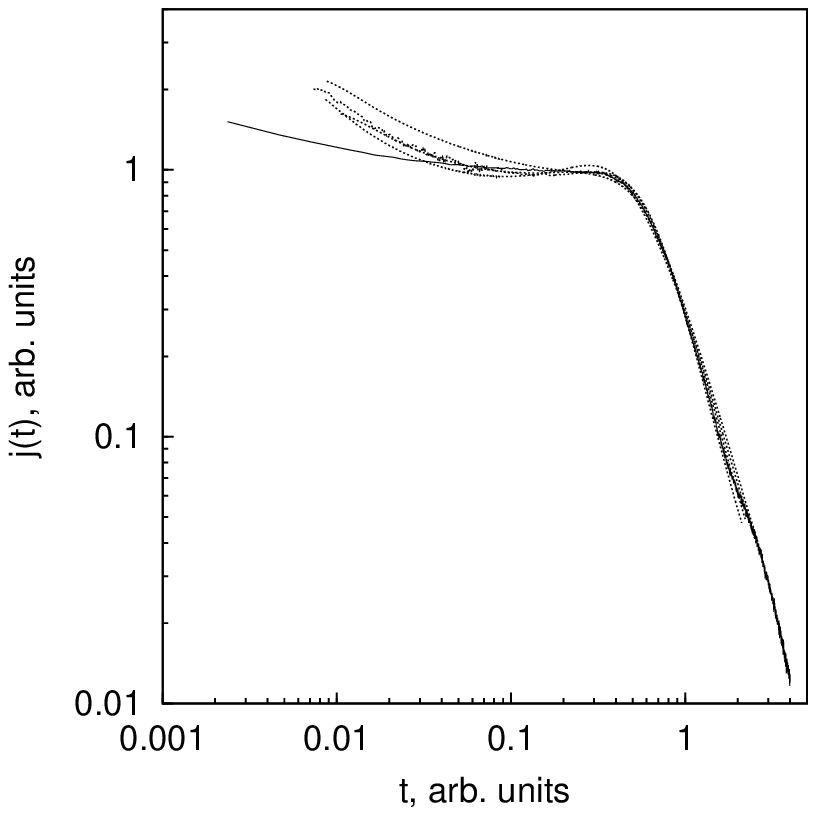}
\end{center}
\caption{Comparison of the simulated TOF transient (solid line) for $L=20\thinspace000$ lattice planes (15.4 $\mu$m) and $F=36$ V/$\mu$m with experimental transients measured at $F=40$ V/$\mu$m in four samples with thickness varied in the range $14-17.5$ V/$\mu$m (dotted lines) \cite{Tyutnev:115107}. Difference in $F$ and $L$ between the simulated transient and experimental ones is unimportant due to universality;  timescales for experimental transients were re-scaled to obtain the best fit (hence, in this figure we compare only the shapes of transients).\label{fig10}}
\end{figure}

Also, the simulated  TOF transients are very similar to experimental ones \cite{Tyutnev:115107}, they demonstrate the same flat plateau and slow power law decay (\fref{fig10}). Systematic deviation at small $t$ is probably connected with electrode effects (roughness of the electrode surface, presence of impurities, variation of the local structure of the material at the vicinity of the electrode).

\section{Conclusions}

We performed Monte Carlo simulation for the DG model with a partially filled lattice where transport molecules  occupy randomly 30\% of all sites. Analysis of simulated TOF transients confirmed the model's unique properties such as observation of Poole-Frenkel mobility field dependence in a broad range from 5 to 200 V/$\mu$m, development of a flat plateau ($\beta_1\approx 0$) followed by the slow power law current decay ($\beta_2\approx  2.0$) and transient universality with respect to both field and thickness.

It has been shown that due to this universality one may use either lin-lin (using $t_0$  as well as $t_{1/2}$ variant) or log-log plots to investigate the field or thickness dependence of the carrier mobility.

Detailed examination of the TOF transients at strong fields reveals the existence of the initial non-equilibrium relaxation of charge carriers. Equilibration time is typically short in comparison to the drift time, does not depend on the sample thickness and exhibits strong field dependence, even more pronounced than the corresponding field dependence of the mobility.

Temporal behavior of the TOF currents (including TOF-2 and TOF-1a) is well reproduced by a simple "quasi-ballistic" model which assumes a statistic (and time-independent) distribution $f(v)$ of carrier drift velocities. If $f(0) = {\rm const} > 0$ and $f(v)$  drops abruptly at some velocity, then this model fairly well describes most prominent features of the simulated transients, the flat plateau followed by the power law decay $j\propto t^{-2}$.

An attempt to fit DG simulation data to a prototype molecularly doped polymer was successful regarding the  Poole-Frenkel slope (0.32 versus 0.39 ($\mu$m/V)$^{1/2}$), if measured at  moderate electric fields, and shape of the transients. To the best of our knowledge, at the moment there is no alternative transport model capable of reproducing both the experimental mobility field dependence and shape of the transients for some particular organic material in the direct MC simulation using transport parameters ($\sigma$, $\gamma a$, and $c$), extracted from the TOF experiments for the same material.

\section*{Acknowledgement}
SVN is grateful for partial financial support from the RFBR grants 10-03-92005-NNS-a and 11-03-00260-a, and  from the Russian Ministry of Education and Science (state contract 16.523.11.3004).


\newpage

\section*{References}

\begin{thebibliography}{10}

\bibitem{Novikov:14573}
S.V. Novikov, A.V. Vannikov, J. Phys. Chem. 99 (1995)
  14573.
\bibitem{Dunlap:542}
D.H. Dunlap, P.E. Parris, V.M. Kenkre, Phys. Rev. Lett.
  77 (1996) 542.
\bibitem{Novikov:4472}
S.V. Novikov, D.H. Dunlap, V.M. Kenkre, P.E. Parris, A.V. Vannikov, Phys. Rev. Lett. 81 (1998) 4472.
\bibitem{Dunlap:437}
D.H. Dunlap, V.M. Kenkre, P.E. Parris, J. Imaging Sci. Tech. 43 (1999) 437.
\bibitem{Parris:126601}
P.E. Parris, V.M. Kenkre, D.H. Dunlap, Phys. Rev. Lett. 87 (2001) 126601.
\bibitem{Gill:5033}
W.G. Gill, J. Appl. Phys. 43 (1972) 5033.
\bibitem{Abkowitz:817}
M.A. Abkowitz, Phil. Mag. B 65 (1992) 817.
\bibitem{Schein:4287}
L.B. Schein, A. Rosenberg, S.L. Rice, J. Appl. Phys. 60 (1986) 4287.
\bibitem{Schein:686}
L.B. Schein, A. Peled, D.J. Glatz, J. Appl. Phys. 66 (1989) 686.
\bibitem{Borsenberger:5188}
P.M. Borsenberger, J. Appl. Phys. 68 (1990) 5188.
\bibitem{Borsenberger:6263}
P.M. Borsenberger, J. Appl. Phys. 68 (1990) 6263.
\bibitem{Borsenberger:465}
P.M. Borsenberger, M.R. Detty, E.H. Magin, Phys. Status Solidi B 185 (1994) 465.
\bibitem{Bassler:15}
H. B{\"a}ssler, Phys. Status Solidi B 175 (1993) 15.
\bibitem{Borsenberger:5327}
P. Borsenberger, H. B{\"a}ssler, J. Chem. Phys. 95 (1991) 5327.
\bibitem{Borsenberger:185}
P.M. Borsenberger, D.S. Weiss, J. Imaging Sci. Tech. 41 (1997) 185.
\bibitem{Novikov:444}
S.V. Novikov, J. Imaging Sci. Tech. 43 (1999) 444.

\bibitem{Dieckmann:8136}
A. Dieckmann, H. B{\"a}ssler, P.M. Borsenberger, J. Chem. Phys. 99 (1993) 8136.

\bibitem{Novikov:181}
S.V. Novikov, D.H. Dunlap, V.M. Kenkre, SPIE Proc. 3471 (1998) 181.

\bibitem{Novikov:2584}
S.V. Novikov, J. Polym. Sci. B 41 (2003) 2584.

\bibitem{Novikov:954}
S.V. Novikov, Annalen der Physik, 18 (2009) 954.

\bibitem{Schein:1067}
L.B. Schein, V.S. Saenko, E.D. Pozhidaev, A.P. Tyutnev, D.S. Weiss, J. Phys. Chem. C 113 (2009) 1067.
\bibitem{Scher:2455}
H. Scher, E.W. Montroll, Phys. Rev. B 12 (1975) 2455.
\bibitem{Rudenko:209}
A.I. Rudenko, V.I. Arkhipov, Phil. Mag. B 45 (1982) 209.
\bibitem{Tyutnev:115107}
A.P. Tyutnev, V.S. Saenko, E.D. Pozhidaev, V.A. Kolesnikov, J. Phys.: Condens. Matter 21 (2009) 115107.
\bibitem{Tyutnev:215219}
A.P. Tyutnev, R.Sh. Ikhsanov, V.S. Saenko, E.D. Pozhidaev, J. Phys.: Condens. Matter 20 (2008) 215219.
\bibitem{Schein:175}%
L.B. Schein, J.C. Scott, L.T. Pautmeier, R.H. Young, Mol. Cryst. Liq. Cryst. 228 (1993) 175.
\bibitem{Borsenberger:967}%
P. Borsenberger, H. B{\"a}ssler, J. Appl. Phys. 75 (1994) 967.
\bibitem{Kreouzis:235201}
T. Kreouzis, D. Poplavskyy, S.M. Tuladhar, M. Campoy-Quiles, J. Nelson, A.J. Campbell, D.D.C. Bradley, Phys. Rev. B 73 (2006) 235201.
\bibitem{Poplavskyy:415}%
D. Poplavskyy, J. Nelson, D.D. Bradley, Macromol. Symp. 212 (2004) 415.
\bibitem{Rudenko:163}
A.I. Rudenko, V.I. Arkhipov, J. Non-Crystal. Solids 30 (1977) 163.
\bibitem{Rudenko:177}
A.I. Rudenko, V.I. Arkhipov, Phil. Mag. B 45 (1982) 177.
\bibitem{Nikitenko:136210}
V.R. Nikitenko, H. von Seggern, H. B{\"a}ssler, J. Phys.: Condens. Matter 19 (2007) 136210.
\bibitem{Borsenberger:12145}
P.M. Borsenberger, L.T. Pautmeier, and H. B{\"a}ssler, Phys. Rev. B 46 (1992) 12145.
\bibitem{Borsenberger:4289}
P.M. Borsenberger, R. Richert, and H. B{\"a}ssler, Phys. Rev. B 47 (1993) 4289.
\bibitem{Novikov:2532}
S.V. Novikov, A.V. Vannikov, J. Phys. Chem. C 113 (2009) 2532.
\bibitem{Mack:7500}
J.X. Mack, L.B. Schein, A. Peled, Phys. Rev. B 39 (1989) 7500.

\end{thebibliography}
\end{document}